%% file: template.tex
\documentclass{Interspeech}

\interspeechcameraready

\title{RELATE: Subjective evaluation dataset for automatic evaluation\\ of relevance between text and audio
}
\author[affiliation={1}]{Yusuke}{Kanamori}
\author[affiliation={1}]{Yuki}{Okamoto}
\author[affiliation={1}]{Taisei}{Takano}
\author[affiliation={2,1}]{Shinnosuke}{Takamichi}
\author[affiliation={1}]{Yuki}{Saito}
\author[affiliation={1}]{Hiroshi}{Saruwatari}

\affiliation{}{The University of Tokyo}{Japan}
\affiliation{}{Keio University}{Japan}
\email{kanamori-yusuke796@g.ecc.u-tokyo.ac.jp, y-okamoto@ieee.org}

\keywords{text-to-audio, human evaluation, CLAPScore, environmental sound synthesis}

\usepackage{amsmath,amssymb}
\usepackage{mathtools}
\usepackage{comment}
\usepackage{bm}
\usepackage{multirow}
\newcommand{\bmit}[1]{{\mbox{\boldmath $#1$}}}
\usepackage{setspace}

\newlength\savedwidth
\newcommand{\wcline}[1]{\noalign{\global\savedwidth\arrayrulewidth\global\arrayrulewidth 1.0pt} \cline{#1}
\noalign{\global\arrayrulewidth\savedwidth}}

\newcommand{\datasetname}{RELATE}
\newcommand{\soundname}{original }

\begin{document}

\maketitle

\input{abstract}

\input{section1}
\input{section2}
\input{section3}
\input{section4}
\input{section5}
\input{section6}
\input{section7}

\bibliographystyle{IEEEtran}
\bibliography{mybib}

\end{document}

%% file: abstract.tex
\begin{abstract}
    
    In text-to-audio (TTA) research, the relevance between input text and output audio is an important evaluation aspect.
    Traditionally, it has been evaluated from both subjective and objective perspectives.
    However, subjective evaluation is costly in terms of money and time, and objective evaluation is unclear regarding the correlation to subjective evaluation scores.
    In this study, we construct {\it RELATE,} an open-sourced dataset that subjectively evaluates the relevance.
    Also, we benchmark a model for automatically predicting the subjective evaluation score from synthesized audio.
    Our model outperforms a conventional CLAPScore model, and that trend extends to many sound categories.
\end{abstract}

    %However, subjective evaluation is costly in terms of money and time, and objective evaluation is reported to have a low correlation with subjective evaluation.

%% file: section1.tex
\vspace{-1mm}
\section{Introduction} \vspace{-1mm}

%Environmental sound synthesis using deep learning has been proposed.
Research on text-to-audio (TTA), which is a technology to automatically synthesize an audio sample from text, such as ``a dog barking behind a human speech,'' is attracting attention \cite{Yang_TASLP_2023}.
%TTA is effective for enhancing immersion and realism in media content such as movies.
TTA has much potential, such as generating background sounds and sound effects for media contents \cite{lloyd2011sound} and creating audio environments in virtual reality.

TTA is evaluated from both subjective and objective perspectives.
%As with speech synthesis, subjective evaluation is a dominant evaluation method for TTA.
Subjective evaluation of TTA can be broadly divided into audio quality and relevance.
The former is an evaluation of whether the synthesized audio samples are of high quality, while the latter is of assessment of the extent to which the synthesized audio samples reflect the content of the input text.
Recent TTA models synthesize high-quality audio samples but often omit content from the input text \cite{Lee_NeuRIPS2024_01}. 
Therefore, we focus on the subjective evaluation of relevance.
%In fact, it is carried out in conjunction with objective evaluation in the DCASE 2024 Challenge Task 7\footnote{\url{https://dcase.community/challenge2024/task-sound-scene-synthesis}} \cite{lagrange2025sound}, international competitions for TTA. 
% [高道] 主観評価軸は大きく分けて，自然性と関連性に分けられる．前者は合成音そのものが自然な音であるかを評価し，後者は合成音が入力テキストの内容をどの程度反映しているかである．昨今の TTA models は，非常に自然な音だが，入力テキストの内容を欠落させた音を合成する~\cite{xxx}．そこで本研究では，relevance の主観評価に着目する．
On the other hand, this requires time and money, and it is impossible to compare scores in different listening tests.
Therefore, realizing an objective evaluation metric that is highly correlated to human subjectivity is an important research topic.

This problem is not limited to TTA but is prevalent in various generative tasks.
To address this issue, supervised machine learning methods have been proposed in speech synthesis and image generation to automatically predict subjective evaluation scores from synthesized outputs \cite{huang2024voicemos,Kou_ACMM2024_01}.
These methods train a machine learning model from paired data of synthesized outputs and subjective evaluation scores, enabling the prediction of subjective evaluation scores for unseen synthesized outputs.
This approach holds promise for application in TTA to simplify evaluation. Furthermore, the subjective evaluation scores of synthesized audio tend to exhibit significant variance among evaluators and synthesized outputs \cite{Okamoto_ICASSP2024_01,Lee_NeuRIPS2024_01}.
When predicting subjective evaluation scores for TTA, it is necessary to investigate and consider the influence of listener, audio, and text attributes.

\begin{figure}[t]
\centering
\includegraphics[width=0.98\linewidth]{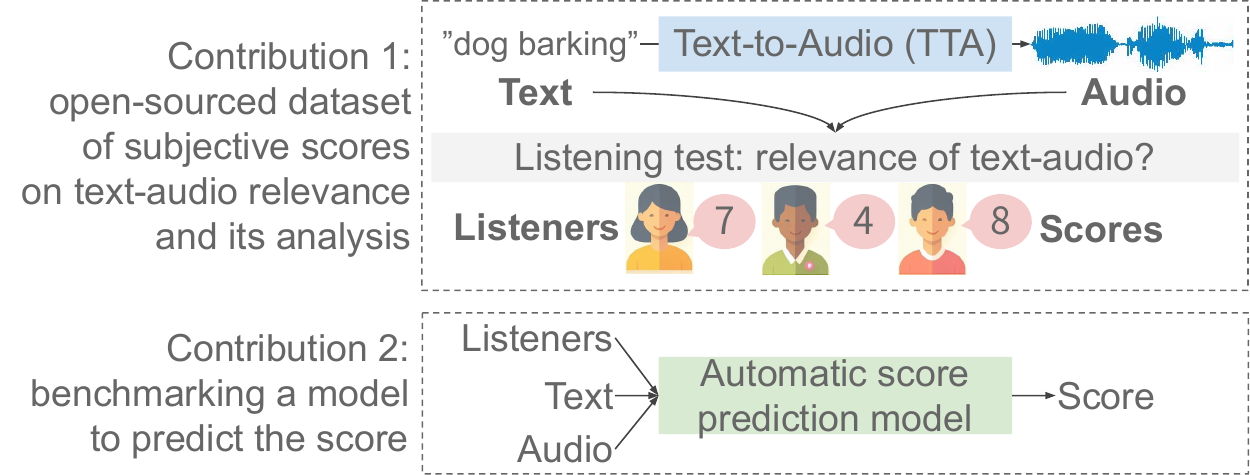}
\vspace{-3mm}
\caption{Overview of this study.}
\vspace{-4mm}
\label{fig:overview}
\end{figure}

The contributions of our paper are illustrated in Figure~\ref{fig:overview}.
We construct an open-source dataset, which is called \textit{\datasetname}~(\textbf{REL}evance score on \textbf{A}udio and \textbf{TE}xt), consisting of synthesized audio samples and relevance scores\footnote{https://github.com/sarulab-speech/RELATE}.
%Table \ref{table:statistics_REL} shows the statistics of XXX dataset.
The collection of scores for synthesized and \soundname audio samples can be expected to be used as a screening method when large amounts of data for TTA are obtained from the internet.
The dataset covers three attributes regarding 1) listener, 2) synthesized audio, and 3) text, and in this study, we investigate the influence of these attributes on subjective evaluation scores.
Furthermore, we train a model using the constructed dataset to predict the relevance between text and audio samples and conduct benchmark analysis.
The results show that our model outperforms CLAPScore \cite{pmlr-v202-huang23i}, and that trend extends to many sound categories.

%% file: section2.tex
\vspace{-1mm}
\section{Related work} \vspace{-1mm}
In the field of speech synthesis, a method for predicting subjective scores on the naturalness of synthesized speech has been proposed.
The VoiceMOS Challenge \cite{huang2024voicemos}, which is an international competition to assess the performance of automatic subjective score prediction, has also been held with a shared dataset of synthesized speech and subjective scores.
Some prediction models have been proposed \cite{saeki22c_interspeech}, and the self-supervised learning (SSL) models (e.g., wav2vec 2.0 \cite{baevski2020wav2vec}, WavLM \cite{Chen_IEEE_2022}) are often used as a module to extract speech features for the subjective score prediction.
In this study, we make use of these ideas and construct a shared dataset for TTA, as well as build an SSL-based benchmark model. Furthermore, we analyze the factors contributing to the variance in subjective evaluation scores.

In the evaluation of TTA, some objective evaluation metrics have been proposed for both overall audio quality (OVL) and relevance to the text input (REL), and their correlation with subjective evaluation scores has been investigated.
Regarding OVL, Deshmukh et al. proposed a prediction method using CLAP \cite{deshmukh24b_interspeech}.
Similarly, Tjandra et al.'s method predicted the values of subdivided evaluation axes of OVL \footnote{A. Tjandra, Y.-C. Wu, B. Guo, J. Hoffman, B. Ellis, A. Vyas, B. Shi, S. Chen, M. Le, N. Zacharov, and others, ``Meta Audiobox Aesthetics: Unified Automatic Quality Assessment for Speech, Music, and Sound,'' arXiv preprint arXiv:2502.05139, 2025.}.
While these studies can predict evaluation scores that correlate to some extent with the subjective evaluation scores of OVL, they cannot be used for the evaluation of REL, which indicates how well the content of the input text is reflected.

Regarding REL, Huang et al. proposed an unsupervised method called CLAPScore \cite{pmlr-v202-huang23i}.
It calculates the cosine similarity between the input text and the synthesized audio using a pretrained CLAP model.
It is unclear what the correlation between the CLAPScore and the REL subjective evaluation scores.
Although there is a difference between unsupervised and supervised learning, this method should be compared with our constructed benchmark.

\begin{comment}
In evaluating TTA, efforts are also being made to achieve objective evaluation that correlates with human subjective evaluation.
Deshmukh et al. have proposed a method for predicting the Overall Quality (OVL) score of synthesized sounds using CLAP \cite{deshmukh2024pam}.
The correlation between the predicted score and the subjective evaluation score for sounds synthesized using the TTA model is currently not high enough, at around 0.6.

In addition to OVL, Tjandra et al. have proposed a method for predicting subjective evaluation scores for the production complexity, content enjoyment, and content usefulness of synthesized sounds \cite{tjandra2025meta}.
This method can be used to evaluate not only the quality of synthesized sounds but also their potential for use in content works.

However, conventional evaluation methods focus only on the quality of the synthesized sound and do not evaluate the output's relevance to the synthesis model's input. In particular, in TTA, the text content used as input reflected in the synthesized sound is a critical evaluation metric. Therefore, we will construct a dataset that evaluates the subjective relevance score of sounds and text to realize a technology that can predict scores highly correlated with human subjective evaluation.
\end{comment}

%% file: section3.tex
\input{listener_attribute}
\vspace{-1mm}
\section{Creation of dataset}\vspace{-1mm}
\subsection{Overview of dataset}\vspace{-1mm}
Our dataset consists of the following contents.

\begin{itemize} \leftskip -0mm \itemsep -1mm
    \item \textbf{Text--audio pairs. }%\\ 
    Both \soundname and synthetic audio samples are included.
    \item \textbf{Subjective evaluation scores. } %
    Three metrics of 11-point scores for each text--audio referring to the DCASE 2024 Challenge Task 7\footnote{\url{https://dcase.community/challenge2024/task-sound-scene-synthesis}}.
        \begin{itemize}
            \item \textbf{REL score. }%\\
            The overall relevance of the text and audio.
            \item \textbf{Inclusion of sound event (IS) score. }%\\
            The extent to which the sound events described in the text are included in the audio.
            \item \textbf{Order of sound event (OS) score. }%\\
            The degree of matching between the time series of sound events described in the text and the audio.
        \end{itemize}
    
    \item \textbf{Listener attributes. }%\\
    Listener ID, age, gender, nationality, birthplace, residence, and experience of audio evaluation.
    Table~\ref{table:experiment_questions} shows the questions and options of listener attributes.
    %The IDs of the listeners who scored each audio-text pair.
\end{itemize}

\input{score_instruction}
\input{dataset_statistics}

\subsection{Collecting subjective evaluation scores}\vspace{-1mm}
We collected subjective scores for each text--audio sample.
%For natural sounds, 1,000 pairs were selected from each of the train and test data of AudioCaps \cite{kim2019audiocaps}.
For \soundname audio samples, 1,000 pairs, including 500 pairs with words indicating the order of sound occurrence, i.e., ``before,'' ``after,'' `then,'' or ``followed by,'' selected on the basis of the conventional study \cite{wu2023audio} and 500 pairs without such words, were randomly selected from each of the AudioCaps training and test datasets \cite{kim2019audiocaps}.
% For natural sounds, 1,000 pairs were randomly selected from each of the AudioCaps train and test datasets  \cite{kim2019audiocaps}.
% This selection included 500 pairs with words indicating the order of sound occurrence (e.g., ``before,'' ``after,'' ``then,'' or ``followed by''), based on conventional studies \cite{wu2023audio}, and 500 pairs without such words.
Synthesized audio samples were obtained using the open-sourced pretrained TTA models: AudioLDM \cite{liu2023audioldm}, AudioLDM2 \cite{liu2024audioldm}, Tango \cite{ghosal2023text}, and Tango2 \cite{majumder2024tango}.
The texts of \soundname audio samples are used as input to these TTA models. For each text, two synthesis models are selected and synthesized.
%\colorbox{green}{The ratio of the amount of data for \soundname and synthetic} \colorbox{green}{audio samples is about 1:2.}
%Referring to the DCASE 2024 Challenge Task 7, the ratings were done on an 11-point scale. 
Furthermore, referring to \cite{otani2023toward}, we presented explanations for each score shown in Table~\ref{table:instruction} to minimize score variations among listeners.
Listeners were presented with the audio and text, and they answered each metric on an 11-point scale.

Each listener answered each metric after being presented with a text--audio pair.
Note that the REL score was obtained through an independent experiment, separate from the IS and OS score collection described later.
In collecting IS and OS scores, the same listeners conducted both evaluations.

In annotating IS scores, some text samples included words indicating the order of sound events, such as ``before'' and ``after.''
Listeners were instructed to disregard the order of occurrence as long as the sound events described in the text were present in the audio.
% In case, in evaluating OS scores, pairs of text and sound that do not represent the order of sound occurrence, listeners were instructed to assign a score of  ``0.''
In annotating the OS score, listeners were instructed to score a ``0'' if the text did not represent the order of sound occurrence and these scores were subsequently omitted from our OS dataset.

\input{analysis_from_audioset}

\subsection{Screening}\vspace{-1mm}
To conduct screening to ensure data quality and the result of evaluation, each evaluation set was intentionally designed to include samples with low relevance between text and audio.
Original audio in AudioCaps is annotated with multiple sound event labels in addition to the text describing the audio.
Audio samples with mismatched sound event labels were randomly selected from the dataset, and these audio samples and texts were paired and set as anchors.

Based on the answers for the anchors in the evaluation, we screened listeners for training and test sets.
For the training set, we excluded listeners whose average anchor rating was ``2'' or higher.
For the test set, we excluded listeners whose average anchor rating was ``1'' or higher to ensure high quality of collected scores.
Table~\ref{table:statistics_REL} shows the statistics of our dataset after the screening.

%% file: listener_attribute.tex
\begin{table}[t]
\centering
\caption{Questionnaire for listener attributes.}
\vspace{-3mm}
\resizebox{\linewidth}{!}{
%\tabcolsep = 3pt
%\footnotesize
\begin{tabular}{l|p{0.64\linewidth}|p{0.36\linewidth}}
ID & Question & Options \\
\hline
Q01 & Age 
    & $\leq$20, 21--30, \ldots, 61$\leq$ \\
Q02 & Gender 
    & M, F, NBi \\
Q03 & How many times have you participated in ratings of audio samples? 
    & 0, 1, \ldots, 5$\leq$\\
Q04 & When did you last participate in other ratings of audio samples? 
    & Never, $\leq$1 month, \ldots \\
Q05 & On average, how many times have you heard an audio repeatedly? 
    & 1, 2, \ldots, 5$\leq$\\
Q06 & What type of audio device did you use? 
    & Headphone, Earphone, Others\\
Q07 & Was the surrounding environment quiet during the ratings of audio samples? 
    & Quiet, \ldots, Noisy \\
Q08 & How difficult were the evaluations? 
    & Easy, \ldots, Difficult\\
Q09 & Do/did you work in the field of speech or audio technology?
    & Yes, No\\
Q10 & Nationality
    & EU, NA, ...\\
Q11 & Mother country
    & EU, NA, ...\\
Q12 & Place of residence
    & EU, NA, ...\\
\end{tabular}
}
\label{table:experiment_questions}
\vspace{-3mm}
\end{table}

%% file: score_instruction.tex
\begin{table}[t]
%\vspace{-5pt}
%\small
\caption{Explanations of each score in subjective evaluations.} 
\label{table:instruction}
\centering
\vspace{-3mm}
%\footnotesize
\resizebox{1.0\linewidth}{!}{%
\begin{tabular}{c|r|l}
    \textbf{Metric} & \textbf{Score} & \textbf{Instruction} \\ \hline
    \multirow{5}{*}{REL}  
    & 0   & Does not match at all. \\
    & 2   & Has significant discrepancies. \\
    & 5   & Has several minor discrepancies. \\
    & 8   & Has a few minor discrepancies. \\
    & 10  & Matched exactly. \\ \hline
    \multirow{5}{*}{IS}  
    & 0 & All sound events are clearly missing. \\
    & 2 & Most of the sound events seem to be missing. \\
    & 5 & About half of the sound events seem to be missing. \\
    & 8 & Most of the sound events seem to be included. \\
    & 10 & All sound events are clearly included. \\ \hline
    \multirow{5}{*}{OS}   
    & 0 & All sound events in the audio clearly occurred in the wrong order.\\
    & 2 & Most sound events in the audio occurred in the wrong order.\\
    & 5 & About half of the sound events in the audio occurred in the correct order. \\ 
    & 8 & Most sound events in the audio occurred in the correct order. \\ 
    & 10 & All sound events in the audio clearly occurred in the correct order. \\
\end{tabular}
 }
 \vspace{-3mm}
\end{table}

%% file: dataset_statistics.tex
\begin{table}[t!]
%\vspace{-5pt}
%\small
\caption{Statistics of RELATE dataset.}
\label{table:statistics_REL}
\vspace{-3mm}
\centering
%\small
\resizebox{\linewidth}{!}{
\begin{tabular}{@{}l|rr|rr|rr@{}}
    \wcline{1-7}
    \\[-9pt]
    & \multicolumn{2}{c|}{{\bf REL}} & 
    \multicolumn{2}{c|}{{\bf IS}} & \multicolumn{2}{c}{{\bf OS}}\\
    & Train & Test  & Train & Test  & Train & Test\\
    \wcline{1-7}
    \\[-9pt]
    Evaluations & 9,963 & 7,797 & 7,641 & 5,865 & 4,017 & 2,943\\
    Audio--text pairs& 2,862 & 2,598 & 2,649 & 2,334 & 1,281 & 1,185\\
    Audio duration [s] & 28,806 & 26,129 & 26,654 & 23,476 & 12,880 & 11901\\
    Listeners & 1,085 & 873 & 864 & 635 & 714 & 525\\
    \wcline{1-7}
\end{tabular}
}
\vspace{-3mm}
\end{table}

%% file: analysis_from_audioset.tex
% \begin{table}[t!]
% %\vspace{-5pt}
% %\small
% \caption{Statistically significant
% differences and interaction with natural/synthetic sounds per factors} 
% \label{table:result_audiolabel}
% \centering
% \vspace{-3mm}
% %\footnotesize
% \resizebox{\linewidth}{!}{%
% \begin{tabular}{@{}lrr@{}}
%     \wcline{1-3}
%     \\[-7pt]
%     {\bf factor} & {\bf significant} & {\bf interaction}\\
%     {} & {\bf difference} & \\
%     \wcline{1-3}
%     \\[-7pt]
%     number of labels (1)&  & \\
%     \cline{1-3}
%     number of Top categories (2)& \checkmark &  \checkmark\\
%     \cline{1-3}
%     Human sounds vs others (3)& & \checkmark \\
%     \cline{1-3}
%     Animal vs others (4)& \checkmark & \checkmark\\
%     \cline{1-3}
%     Natural sounds vs others (5)& & \checkmark\\
%     \cline{1-3}
%     Music vs others (6)& \checkmark & \\
%     \cline{1-3}
%     Sounds of things vs others (7)& & \\
%     \cline{1-3}
%     Source-ambiguous sounds vs others (8)& & \\
%     \cline{1-3}
%     Channel, environment & & \\
%      and background vs others (9)& & \\
%      \cline{1-3}
%     Speech vs others (10)& & \checkmark \\
%     \wcline{1-3}
%     \\[-7pt]
%     number of words (11) & \checkmark & \checkmark\\
%      \cline{1-3}
%     w/ temporal words vs others (12)& \checkmark &  \checkmark\\
%      \cline{1-3}
%     Flesch reading ease (13) & \checkmark & \\
%     \wcline{1-3}
% \end{tabular}
%  }
%  \vspace{-3mm}
% \end{table}

\begin{table}[t]
%\vspace{-5pt}
%\small
\caption{Statistical significances ($p < 0.05$) among items and of interaction between items of natural/synthetic audio samples for REL scores. Check mark shows significant differences and interaction.} 
\label{table:result_audiolabel}
\centering
\vspace{-3mm}
%\footnotesize
\resizebox{\linewidth}{!}{%
\begin{tabular}{l|cc}
    \textbf{Factor}                     & \textbf{Among items}   & \textbf{Interaction} \\ \hline
    \multicolumn{3}{l}{Sound event labels in text}              \\ \hline
    Number of event labels              &               & \\
    Number of top-level sound categories      & \checkmark    & \checkmark\\ \hline

    \multicolumn{3}{l}{Sounds belonging to a category vs. sounds in other categories}              \\ \hline
    Human sounds             &               & \checkmark\\
    Animal                    & \checkmark    & \checkmark\\
    Natural sounds            &               & \checkmark\\
    Music                     & \checkmark    & \\
    Sounds of things          &               & \\
    Source-ambiguous sounds   &               & \\
    Channel, environment and background&      & \\
    Speech                    &               & \checkmark \\ \hline

    \multicolumn{3}{l}{Text complexity}              \\ \hline
    Number of words                     & \checkmark    & \checkmark\\
    w/ temporal preposition vs. w/o     & \checkmark    &  \checkmark\\
    Flesch Reading Ease                 & \checkmark    & \\
\end{tabular}
 }
 \vspace{-3mm}
\end{table}

%% file: section4.tex
\begin{figure}[t!]
  \centering
  \includegraphics[scale=1.0]{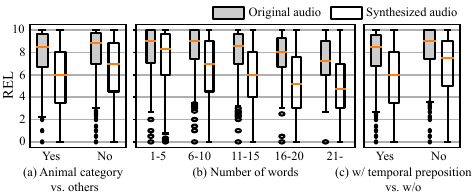}
  \vspace{-5mm}
  \caption{Boxplots of each aspect}
  \label{figure:boxplot}
\end{figure}
\begin{figure}[t]
\centering
\includegraphics[width=0.98\linewidth]{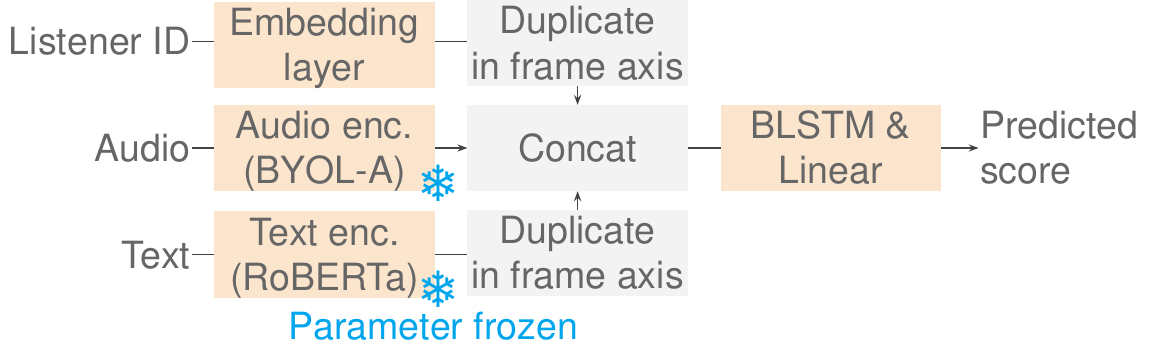}
\vspace{-3mm}
\caption{Overview of proposed REL prediction model}
\label{fig:model}
\end{figure}
\input{training_statistics}
\input{prediction_result}
\input{SRCC_each_event_category}

\vspace{-1mm}
\section{Analysis of dataset} \vspace{-1mm}
\label{analysis_of_dataset}
REL is the most commonly used metric in three metrics, and we have conducted an analysis of REL.
We focused on the REL and analyzed the subjective evaluation dataset in the following aspects: audio attribute and text attribute.
Furthermore, we investigated whether these trends differ between \soundname and synthesized audio samples.
A stricter screening than those in the dataset creation process was performed in this analysis to reduce noise in the evaluation and to acquire meaningful tendency.
In addition to excluding listeners with an average anchor rating of ``2''  or higher, we added the exclusion of listeners with an average \soundname audio rating of ``6'' or lower to avoid listeners who rated everything low. Then, we removed listeners whose ratings had the lowest 5\% entropy of all listeners.

We conducted nonparametric tests: Mann--Whitney U test \cite{mann1947test} for two-group means, Kruskal--Wallis test \cite{kruskal1952use} for 3+-group means, and Steel--Dwass test \cite{steel1959multiple} for multiple comparison.
In addition, aligned rank transform (ART) analysis of variance (ANOVA) \cite{wobbrock2011aligned}, a method for nonparametric data,  was used to examine the interactions of two factors.

In TTA, audio attributes and text attributes are important.
Original audio samples of AudioCaps have labels indicating the type of sound.
This label is reflected in the text, which in turn is reflected in the synthesized audio.
Then, those labels can be applied to synthesized audio samples as audio attributes. 
Text attributes are characterized by their complexity. We examined our dataset using audio attributes and text attributes in the following three aspects.
\begin{itemize}
    \item \textbf{Sound event labels.}  
    The number of sound event labels and the number of top-level categories.
    \item \textbf{Sounds belonging to a category vs. sounds in other categories.}
    Sounds belong to a top-level category and others, e.g., human sounds vs. others.
    
    \item \textbf{Text complexity.} 
    Number of words in the text, with or without temporal prepositions, and the Flesch Reading Ease score \cite{flesch1948new}, which indicates the readability of English text.
\end{itemize}

\vspace{-1mm}
\subsection{Analysis from sound event label}\vspace{-1mm}
\label{sec:analysis_from_event_label}
Upper part of Table~\ref{table:result_audiolabel} shows which audio factors had statistically significant differences and which audio factors interacted with original/synthetic audio samples. 
It shows that top-level categories affect the evaluation score. It is assumed that the kind of top-level categories affects both the difficulty of human evaluation and that of synthesis. Specifically, belonging to the ``Animal'' category shows both of statistically significant differences and interaction. Figure~\ref{figure:boxplot}(a). shows that synthesized audio in the ``Animal'' category had the lower score than in others. It is assumed that synthetic models have difficulty in synthesizing animal sounds.

%%%%%%
\vspace{-1mm}
\subsection{Analysis from texts}\vspace{-1mm}
\label{analysis_from_caption}
Lower part of table \ref{table:result_audiolabel} shows which text factors had statistically significant differences and which text factors interacted with original/synthetic audio samples. In particular, number of words and inclusion of temporal preposition show both of statistically significant differences and interaction. Figure~\ref{figure:boxplot}(b) shows that the score decreases as the number of words increases, and interaction effects indicate that the more words there are, the less successful the synthesis is. This can be thought of as the greater the number of words, the more difficult the subjective evaluation or synthesis becomes.
Figure~\ref{figure:boxplot}(c) shows that
the score decreases for texts that include time-series information.
The interaction effects indicate that the synthesis is not successful when time-series information is included.
It is inferred that the synthetic models are not trained with attention to time-series information.

%% file: training_statistics.tex
\begin{table}[t!]
%\vspace{-5pt}
%\small
\caption{Distribution of training, validation, and test data.}
\label{table:training_statistics}
\vspace{-3mm}
\centering
%\small
\resizebox{0.70\linewidth}{!}{
\begin{tabular}{@{}l|r|rr@{}}
    %\wcline{1-4}
    \hline
    \\[-9pt]
    %& \multicolumn{3}{c}{{\bf REL}} \\
    %from XXX dataset & Train &\multicolumn{2}{c} {Test}\\
    %\hline
    AudioCaps & Train &\multicolumn{2}{c} {Test}\\
    %\cline{2-4}
    \hline
    RELATE dataset & Train & Validation  & Test \\
    %\wcline{1-4}
    \hline
    \\[-9pt]
    Evaluations & 9,963 & 3,897 & 3,900 \\
    Audio--text pairs& 2,862 & 1,287 & 1,311 \\
    Audio duration [s] & 28,806 & 12,960 & 13,169 \\
    Listeners & 1,085 & 712 & 726 \\
    %\wcline{1-4}
    \hline
\end{tabular}
}
\vspace{-3mm}
\end{table}

%% file: prediction_result.tex
% \begin{table}[t!]
% %\vspace{-5pt}
% %\small
% \caption{Results of subjective evaluation prediction}
% \label{table:result_predict}
% \centering
% \small
% \begin{tabular}{@{}lrrrr@{}}
%     \wcline{1-5}
%     \\[-9pt]
%     Method & MSE $\downarrow$& LCC $\uparrow$& SRCC $\uparrow$& KTAU $\uparrow$\\ 
%     \wcline{1-5}
%     \\[-9pt]
%     MS-CLAP & $0.159$ & $0.208$ & $0.181$ & $0.124$\\
%     LAION-CLAP & $0.082$ & $0.375$ & $0.351$ & $0.242$\\
%     Ours w/o CBL& $\bm{0.069}$ & $0.377$ & $0.374$ & $0.259$\\
%     Ours w/ CBL & $0.073$ & $\bm{0.385}$ & $\bm{0.383}$ & $\bm{0.265}$\\
%     \wcline{1-5}
% \end{tabular}
% \end{table}

\begin{table}[t]
\small
\caption{Results of subjective evaluation prediction.}
\vspace{-3mm}
\label{table:result_predict}
\centering
\resizebox{\linewidth}{!}{
\begin{tabular}{l|rrrr}
    Method                  & MSE $\downarrow$& LCC $\uparrow$& SRCC $\uparrow$& KTAU $\uparrow$\\ \hline
    CLAPScore w/ MS-CLAP    & $0.159$ & $0.208$ & $0.181$ & $0.124$\\
    CLAPScore w/ LAION-CLAP & $0.082$ & $0.375$ & $0.351$ & $0.242$\\
    Ours                    & $0.073$ & $\textbf{0.385}$ & $\textbf{0.383}$ & $\textbf{0.265}$\\
    Ours w/o CBL            & $\textbf{0.069}$ & $0.377$ & $0.374$ & $0.259$\\
\end{tabular}
}
\vspace{-3mm}
\end{table}

%% file: SRCC_each_event_category.tex
\begin{table*}[t]
\small
\caption{Results of subjective evaluation prediction for each top-level category.}
\vspace{-3mm}
\label{table:result_predict_each_event}
\centering
\resizebox{0.80\linewidth}{!}{
\begin{tabular}{l|rrrrrrrr}
     & &  &  &  &  & Source- & Channel, & \\
    \multirow{2}{*}{Method}  & Human& \multirow{2}{*}{Animal} & Natural & \multirow{2}{*}{Music} & Sounds & ambiguous & environment & \multirow{2}{*}{Speech}\\
    &sounds&&sounds&&of things&sounds&and background&\\ \hline
    CLAPScore w/ MS-CLAP  & $0.181$ & $0.254$ & $0.232$ & $0.310$ & $0.126$ & $0.339$ & $0.170$ & $0.174$\\
    CLAPScore w/ LAION-CLAP & $0.349$ & $\textbf{0.438}$ & $0.311$ & $0.186$ & $0.266$ & $\textbf{0.452}$ & $0.338$ & $0.339$\\
    Ours                    & $\textbf{0.435}$ & $0.411$ & $\textbf{0.353}$ & $\textbf{0.599}$ & $\textbf{0.380}$ & $0.447$ & $\textbf{0.462}$ & $\textbf{0.445}$\\
\end{tabular}
}
\vspace{-3mm}
\end{table*}

\begin{comment}
\begin{table*}[h]
  \begin{minipage}[t]{0.05\columnwidth}
    %\begin{center}
      \begin{tabular}{l|rrrr}
    Method                  & MSE $\downarrow$& LCC $\uparrow$& SRCC $\uparrow$& KTAU $\uparrow$\\ \hline
    CLAPScore w/ MS-CLAP    & $0.159$ & $0.208$ & $0.181$ & $0.124$\\
    CLAPScore w/ LAION-CLAP & $0.082$ & $0.375$ & $0.351$ & $0.242$\\
    Ours                    & $0.073$ & $\textbf{0.385}$ & $\textbf{0.383}$ & $\textbf{0.265}$\\
    Ours w/o CBL            & $\textbf{0.069}$ & $0.377$ & $0.374$ & $0.259$\\
\end{tabular}
    %\end{center}
  \end{minipage}
  %
  %
  \begin{minipage}[t]{0.05\columnwidth}
    %\begin{center}
      \begin{tabular}{l|rrrrrrrr}
     & &  &  &  &  & Source- & Channel, & \multirow{2}{*}{Speech}\\
    \multirow{2}{*}{Method}  & Human& \multirow{2}{*}{Animal} & Natural & \multirow{2}{*}{Music} & Sounds & ambiguous & environment & \multirow{2}{*}{Speech}\\
    &sounds&&sounds&&of things&sounds&and background&\\ \hline
    CLAPScore     & $0.349$ & $\textbf{0.438}$ & $0.311$ & $0.186$ & $0.266$ & $\textbf{0.452}$ & $0.338$ & $0.339$\\
    CLAPScore  & $0.181$ & $0.254$ & $0.232$ & $0.310$ & $0.126$ & $0.339$ & $0.170$ & $0.174$\\
    Ours                    & $\textbf{0.435}$ & $0.411$ & $\textbf{0.353}$ & $\textbf{0.599}$ & $\textbf{0.380}$ & $0.447$ & $\textbf{0.462}$ & $\textbf{0.445}$\\
\end{tabular}
    %\end{center}
  \end{minipage}
\end{table*}
\end{comment}

%% file: section5.tex
\vspace{-1mm}
\section{Benchmarking prediction model}\vspace{-1mm}
\subsection{Model architecture}\vspace{-1mm}
\label{mode_structure}
We trained a model to predict the REL score between audio and text.
Figure~\ref{fig:model} shows the model.
The audio $\bmit{x}$ and text $w$ are input into the pretrained audio and text encoders, respectively:
$\bm{V}=\mathsf{AudioEnc}\left(\bmit{x}\right)\in\mathbb{R}^{F\times T}, \,     \bmit{o}=\mathsf{TextEnc}\left(w\right)\in\mathbb{R}^{D}.$
Here, $F$, $T$, and $D$ denote the number of dimensions of audio features, time length, and number of dimensions of text features, respectively. 
For audio and text encoders,  we used pre-trained BYOL-A \cite{niizumi2022byol} and RoBERTa\footnote{Y. Liu, ``Roberta: A robustly optimized bert pretraining approach,'' arXiv preprint arXiv:1907.11692, 2019.}, respectively.
Then, $\bmit{o}$ and $C$-dimension listener-embedding vector $\bm{l}\in \mathbb{R}^{C}$ are duplicated in the time direction. 
Listener embeddings enhance prediction accuracy by modeling individual listener preferences \cite{huang2022ldnet}.
The feature $\bm{M}\in\mathbb{R}^{(F+C+D)\times T}$ obtained by concatenating the obtained feature sequences $\bm{V}$and the temporally duplicated ${\bmit l}, {\bmit o}$ in the dimension direction is input to the bidirectional long short--term memory (BLSTM) \cite{graves2005bidirectional}.
Finally, by passing the output $\bm{Z}$ from the BLSTM through two linear layers and the activation function ReLU, the REL score is predicted.

\subsection{Loss function}\vspace{-1mm}
 The training objective is a weighted sum of two functions, both evaluating the difference between the ground truth and predicted scores: the clipped mean squared error (MSE) loss $\mathcal{L}^{reg}$ \cite{leng2021mbnet} and contrastive loss $\mathcal{L}^{con}$ \cite{saeki22c_interspeech}.
Class-balanced loss (CBL) \cite{cui2019class} was introduced to reduce the influence of data bias.
We round up REL scores to the nearest integer. Let $l_y \in [1, 2, ..., 10]$ be the integer-converted version of REL score $y$, and $n_{l_y}$ be the frequency of class $l_y$. Then, $E_y=\frac{1-\beta_{cbl}}{1-\beta_{cbl}^{n_{l_y}}}$ is a value that decreases as $n_{l_y}$ increases. Here, $\beta_{cbl}\in[0,1]$ is a hyperparameter.
Then, $\mathcal{CBL}^{reg}$ and $\mathcal{CBL}^{con}$ are obtained by applying $E_y$ to each loss function.
The final loss function is obtained by summing the two loss functions with weights $\beta$ and $\gamma\colon$ $\mathcal{L}=\beta\mathcal{CBL}^{reg}+\gamma\mathcal{CBL}^{con}$.

\subsection{Experimental setup}\vspace{-1mm}
Table~\ref{table:training_statistics} shows the distribution of training, validation, and test data of REL scores in the RELATE dataset.
For model training, we used the training data of REL scores.
For validation and evaluation, the test data of REL scores was divided into two subsets so there was no overlap between audio samples and texts.
The prediction score was normalized to the range of $[-1,1]$.
In addition to each listener, score prediction was performed for the average listener, whose score is the average of the scores of all listeners \cite{huang2022ldnet}.
During inference, the model predicted the average listener's score.
We empirically chose $\tau=0.25$ for clipped MSE loss, $\alpha=0.1$ for contrastive loss, $\beta_{cbl}=0.99$ for CBL, and $\beta=1.0$ and $\gamma=0.5$
for the weights of the two loss functions.
The batch size was 12, and gradient accumulation was performed every two steps. Adam \cite{Kingma_ICLR2015_01} $(\beta_1 = 0.9, \beta_2 = 0.999)$ was used as the optimizer with an initial learning rate of $2.0 \times 10^{-5}$. Also, learning rate scheduling with linear warm-up and linear decay was used. The total number of training steps was 15,000, with up to 4,000 warm-up steps. The optimal model was selected referring to Spearman's rank correlation coefficient (SRCC) calculated from the validation set.

We used CLAPScore for the comparison.
As the CLAP model used to calculate CLAPScore, LAION-CLAP \cite{wu2023large} and MS-CLAP \cite{Elizalde_ICASSP2023_01} were used.
For the evaluation of the prediction scores of each model, we used MSE, linear correlation coefficient (LCC), SRCC, and Kendall rank correlation coefficient (KTAU) referring to the VoiceMOS Challenge \cite{huang2024voicemos}.

\subsection{Results and discussion}\vspace{-1mm}
Table~\ref{table:result_predict} shows the results for each evaluation metric. The proposed method outperforms both MS-CLAP and LAION-CLAP in all metrics.
It can be seen that the output of the proposed method is closer to the human subjective evaluation value than the similarity score calculated from MS-CLAP and LAION-CLAP.
We also found that CBL is effective.

To capture trends in the strengths and weaknesses of subjective evaluation score prediction, SRCCs of prediction and subjective evaluation values were calculated for each top category and compared with the CLAPScore.
Table~\ref{table:result_predict_each_event} shows the results. Our method outperformed LAION-CLAP and MS-CLAP for all categories except two cases, ``Animal'' and ``Source--ambiguous sounds.''
In particular, a large difference is observed for ``Music,'' which may be because this method, compared with CLAPScore, reflects the tendency of subjective evaluation to recognize and rate highly music, which has characteristics that are very different from those of other sounds.

Further improvements could be made to the model structure.
In the future, experiments using audio encoders other than BYOL-A will be necessary.
The quality of the audio encoder is paramount, especially in the field of TTA, which deals with many types of audio.
Regarding the text encoder, explicitly encoding the units of sound events, rather than encoding the text as it is, may be helpful for REL prediction.
As for the input, we believe that listener attributes help in predicting subjective evaluation scores by providing modeling of the listener.

%% file: section6.tex
\vspace{-1mm}
\section{Conclusion} \vspace{-1mm}
We constructed an open-source dataset consisting of synthesized audio samples and relevance scores.
From the analysis result, regarding audio attributes, people have different evaluation tendencies for different types of sound, and there are sounds that the synthesis model does not handle well.
For text complexity, we found that the longer the sentence, the lower the evaluation value and the less successful the synthesis. Also, synthesis from texts containing time series was poor.
We also built a baseline model for predicting the subjective evaluation score of text--audio relevance in TTA. 
Our model outperformed the conventional method, CLAPScore, and that trend extended to many sound categories.
For future work, we are considering the analysis and prediction of IS and OS scores, the development of a dataset that focuses on audio attributes and text attributes, and the improvement of the prediction model by introducing other encoders or inputs.

%% file: section7.tex
\vspace{-1mm}
\section{Acknowledgements} \vspace{-1mm}
The work was supported by JSPS KAKENHI Grant Number 23K24895, 24K23880, 25K21221, JST Moonshot Grant Number JPMJMS2237.